# Economics in Nouns and Verbs

*By* W. BRIAN ARTHUR [1]

5 April 2021

**Abstract.** Standard economic theory uses mathematics as its main means of understanding, and this brings clarity of reasoning and logical power. But there is a drawback: algebraic mathematics restricts economic modeling to what can be expressed only in quantitative nouns, and this forces theory to leave out matters to do with process, formation, adjustment, and creation—matters to do with nonequilibrium. For these we need a different means of understanding, one that allows verbs as well as nouns. Algorithmic expression is such a means. It allows verbs—processes—as well as nouns—objects and quantities. It allows fuller description in economics, and can include heterogeneity of agents, actions as well as objects, and realistic models of behavior in ill-defined situations. The world that algorithms reveal is action-based as well as object-based, organic, possibly ever-changing, and not fully knowable. But it is strangely and wonderfully alive.

Science proceeds as much by its instruments—its technologies—as it does by human thought.[2] In early January 1610, when Galileo directed his telescope at the heavens he found to his astonishment that Jupiter had four companion "stars" (which after several nights he realized were moons circling Jupiter), and the Moon itself had mountains and valleys. This went against the long-accepted truths that all heavenly bodies circled the earth (or the sun), and that all were perfect. Instruments in science probe, they reveal, they occasionally surprise, and they illuminate. They become means of understanding.

What fascinates me is that different means "see" differently—often they reveal different versions of the same object. Magnetic resonance imaging (MRI) looks at body parts and reveals soft tissue structure; computerized tomography (CT scanning) looks at the same body parts and reveals bone structure. Similarly, when standard light-spectrum telescopes examine the Sun they see a sphere of radiating energy with some surface phenomena; when X-ray telescopes examine the Sun they see a sphere riven by jagged fissures and far from uniform. There is no "correct" version of internal body parts, or of the Sun for that matter. These things exist in reality but there is no correct way to "see" them.

Mathematics I believe is an instrument of understanding, albeit a sublimely mental one. It probes, it reveals, it occasionally surprises, and it illuminates how things follow from the premises we choose. As a method it arrived seriously in economics in the 1870s with the work of Stanley Jevons, Léon Walras, and Francis Edgeworth. And it stirred some degree of contention even then. The economy is very much a creation of humans and a very complicated one, so given the liberty of human choices and the vagaries of people's actions, it is not obvious why algebraic logic and calculus should apply. "Some say human liberty will never allow itself to be cast into equations," said Walras in 1874, speaking of his detractors; then he dismissed them with a shrug, "Those economists who do not know any mathematics, … let them go their way" (quoted in McCloskey, 2005).

---

After a passage of some 150 years, both sentiments are still with us.

I should confess my own position before we go further. I am a theorist in economics and a great deal of my training has been in mathematics. I believe mathematics is powerful in economics—and necessary—but I don't believe that it is suited to describing all that is interesting in an economy. In fact I don't believe that given the complication of the economy and the humanness of the people who act within it, there is any privileged way to view the economy. There are useful ways and less useful ones. In economics we have many means of understanding besides mathematics: narrative discourse based on close observation; insights via geometry, or via computation, or via economic history, or via examining statistics. All these show different things; and all I believe are legitimate.

And so we can well ask in this very human of creations, the economy, what does the instrument of mathematics especially reveal? And what does it work not to reveal? To give a hint, I will argue that economics, as expressed via algebraic mathematics, deals only in nouns—quantifiable nouns—and does not deal with verbs (actions), and that this has deep consequences for what we see in the economy and how we theorize about it.

# I

## The Noun-based Science

Let me begin by pointing out that economics deals with prices, quantities produced, consumption, rates of interest, rates of exchange, rates of inflation, unemployment levels, trade surpluses, GDP, financial assets, Gini coefficients. These are all nouns. In fact, they are all quantifiable nouns—amounts of things, levels of things, rates of things. Economics as it is formally expressed is about amounts and levels and rates, and little else. This statement seems simple, trite almost, but it is only obvious when you pause to notice it. Nouns are the water economics swims in.

Of course in the real economy there are actions. Investors, producers, banks, and consumers act and interact incessantly. They trade, explore, forecast, buy, sell, ponder, adapt, invent, bring new products into being, start companies. And these of course are actions—verbs. Parts of economics—economic history, or business reporting—do deal with actions. But in formal discourse about the economy, in the theory we learn and the models we create and the statistics we report, we deal not with verbs but with nouns. If companies are indeed started, economic models reflect this as the number of companies started. If people invest, models reflect this as the amount of investment. If central banks intervene, they reflect this by the quantity of intervention. Formal economics is about nouns and reduces all activities to nouns. You could say that is its mode of understanding, its vocabulary of expression.

Perhaps this is just a curiosity and doesn't matter. And maybe it's necessary that to be a science economics needs to deal with quantifiable objects—nouns. But other sciences heavily use verbs and actions. In biology DNA replicates itself, corrects copying errors in its strands, splits apart, and transfers information to RNA to express genes. These are verbs, all. Biology—modern molecular biology, genomics, and proteomics—is based squarely on actions. Indeed biology would be hard to imagine without actions—events triggering events, events inhibiting events.

So why then is economics noun-based? One good reason is that any field, be it musicology or jurisprudence or architecture, as it becomes theoretical, tends to categorize and marshal its thinking into concepts and often these concepts are nouns. They are real or abstract things. But this can only be a partial explanation. Concepts don't need to be nouns, they can also be verbs—processes—as is true in biology, so something else must be at work.

That something else, I believe, is that formal economics—theoretical economics—is expressed very largely in algebraic mathematics, a language that allows quantitative nouns only, but doesn't allow actions.[3] So let us take a closer look at that mathematical mode of expression.

---

[3] In 1992 Kenneth Boulding pointed out that "Mathematics is particularly impaired given its deficiency of verbs; these are limited to just four, namely, is equal to, is greater than, is less than, or is a function of."

**Mathematics and what it can express**

In modern times, economic theory has become mathematized to the degree that we often think of theory as mathematics. But theory is not mathematics. Theory lies in the discovery, understanding, and explaining of phenomena present in the world; it is at bottom a set of principled explanations for what we see in the world. And as such it is put together—constructed—always by some language of expression, and there are several such languages for economics. If "reality" is a landscape before us and theory is a painting of this, then a language of expression is the palette used. Different palettes allow us to see different things, and each allows us to express in generalizable terms some truism of what we are looking at.

What then does mathematics allow us to express?

Before we go further let me say I've been talking loosely about "mathematics" when I mean standard algebra, calculus, and linear algebra as used in economics journals or in engineering or physics. I'll use the shorthand labels "algebra" or "algebraic mathematics" or "equation-based mathematics" for these from here on. The full discipline of mathematics of course is much wider and encompasses group theory, topology, number theory, logic, probability theory, and so on. Later I'll be talking about a different form of mathematics, algorithmic mathematics.

How then does this standard equation-based mathematics work?

Consider an arbitrary simple algebraic equation system such as might be studied in high school, and notice how it works.

$$13x + 9y = 790$$
$$5x + 15y = 650$$

The x's and y's here of course are unknown quantities. Algebra at its base is about starting with the "story" of a problem: "A shepherd in Tuscany sells 13 sheep and 9 goats for 790 denari; his son sells 5 sheep and 15 goats … ." We abstract from the story unknown quantities (the prices x and y) and designate them as symbols, then relate these unknowns to known quantities, and after some manipulations that preserve the equalities we can state the unknowns in terms of known quantities.

Notice two things about this.

The first is that algebra is a form of arithmetic: it performs arithmetic operations—addition, subtraction, multiplication, raising to powers, taking square roots—on numerical quantities.[4] (Indeed Isaac Newton called his 1707 treatise on algebra *Universal Arithmetick*.) The numerical quantities in this arithmetic are by definition nouns. Algebra must therefore always deal with nouns.

Given this, it follows that to use algebraic equations a field of interest must be reduced to nouns.[5]

The second thing is that whatever science it is used in, algebra is always a narrative. It tells a story in quantified terms. To do this it sets up a situation with characters (symbols) and their relations to one another, and much like a Sherlock Holmes tale it manipulates these in ways that preserve deductive logic to arrive at some form, some unfolding, that tells us something we want to know. But the story it sets up is not in prose narrative, it is in symbolic and abstracted narrative. To put our story into the equations above, you have to clean it up. The shepherd and his son must disappear, the sheep and goats must be reduced to numbers, the unknown prices must become symbols, and any whiff of haggling or trading or dithering at the market must vanish. To be readied for manipulation, the algebraic version must be stripped of behavior, stripped of action, and separated from any dross of personalities or their inclinations. That way it becomes mathematics.

Algebra goes back at least to the 850s, and for its first 750 years it was an art exercised in wordy form by merchant adepts (Devlin 2017, Maloney 1977). Then in the late 1500s symbols arrived, and natural philosophy took notice. The symbolic notations added a further aura of mystery. But whatever the mystique and the strange glyphs, algebra remained storytelling in a symbolic way.

Algebra, when it got going in the 1600s, brought the distinct advantage of fast logical ma-nipulation of the essentials of a problem. Descartes was able to take a puzzle Euclid had solved in a welter of geometry and knock it off in 5 lines (Gaukroger, 1995, p.175). Yet in spite of such feats, algebra wasn't fully accepted in the sciences throughout the 1600s. At that time geometry was mathematics, and mathematics was geometry. Hobbes in 1656 lambasted the new algebra as "a scab of symbols [which disfigured the page] as if a hen had been scraping there." And Newton wrote to Scottish mathematician David Gregory that "algebra is the analysis of bunglers in mathematics" (Kline 1980). Algebra became accepted as a language of expression in the physical sciences only around 1720; and now of course it has become fundamental.

---

[4] Similarly, calculus, which is an extension of algebra (with its own rules), is again a form of arithmetic.

[5] This is why biology, which can't easily be reduced to nouns, is only partially mathematizable.

§

How does this powerful instrument of thinking work in economics? It certainly offers precision of thought and clarity of expression. But it also brings a different style of expression. Let us compare 2 examples of this.

First the non-mathematical style. Here is Alfred Marshall in his Principles of Economics (1890) explaining agglomeration economies, the tendency for an industry to cluster in a given region:

> When an industry has thus chosen a locality for itself, it is likely to stay there long: so great are the advantages which people following the same skilled trade get from near neighbourhood to one another. The mysteries of the trade become no mysteries; but are as it were in the air, and children learn many of them unconsciously. Good work is rightly appreciated, inventions and improvements in machinery, in processes and the general organization of the business have their merits promptly discussed: if one man starts a new idea, it is taken up by others and combined with suggestions of their own; and thus it becomes the source of further new ideas. And presently subsidiary trades grow up in the neighbourhood, supplying it with implements and materials, organizing its traffic, and in many ways conducing to the economy of its material.

Notice the languid depiction of reality here: we have children learning, ideas taken up, mysteries in the air. Marshall himself was an excellent mathematical economist, yet he tells the story in words. There are nouns: advantages, inventions, improvements, suggestions. And verbs: industry chooses, good work is appreciated, inventions are discussed, subsidiary trades grow up, supply implements, organize its traffic. Marshall conjures the notion of "agglomeration economies" alive by invoking the real world actions that create them.

Here by contrast, is Paul Samuelson 60 years later, using full-on algebraic mathematics to explain international trade. Samuelson imagines a scientist doing economic planning:

> Here is the question the scientist has posed for himself. He wants (a) to maximize the raw total of clothing production in Portugal and England, subject to (b) a prescribed total for food, and subject to (c) the two linear production possibility constraints of the two countries, where (d) all quantities must by their nature not be negative numbers. Mathematically,
>
> $Z = X_2 = x_2 + x_2'$ is to be maximum subject to
>
> (4)
> $$x_1 + x_1' = X_1$$
> $$2x_1 + x_2 \leq C$$
> $$x_1' + x_2' \leq C'$$
> $$\text{and } x_1 \geq 0,\ x_2 \geq 0,$$
> $$x_1' \geq 0,\ x_2' \geq 0$$

We needn't worry about the interpretation here; but notice the objects of discussion. They are symbolic quantities expressed in relation to other symbolic quantities—nouns related to other nouns. Any wordy narrative has been stripped; and any people trading, like the Tuscan shepherds, have vanished. There is only the faceless scientist. There are actions behind the nouns presumably, but only presumably; they are now invisible and unspecified—the language does not allow them. And of course no such scientist exists, optimally directing trade between England and Portugal. The problem has been fitted to the technique, here simple linear programming. We can debate the usefulness of Samuelson's language of explanation. What I want to point to here is the lack of verbs. Verbs have vanished.

The reader may object that mathematics in economics does use verbs: agents maximize; they learn and adapt; rank preferences; decide among alternatives; adjust supply to meet demand. But the verbs here are an illusion; algebraic mathematics doesn't allow them, so they are quickly finessed into noun quantities. We don't actually see agents maximizing in the theory; we see the necessary conditions of their maximizing expressed as noun-equations. We don't see them learning; we see some parameter changing in value as they update their beliefs. We don't see producers deciding levels of production; we see quantities determined via some optimizing rule. We don't see producers responding to demand via manufacturing actions, we see quantities adjusting. It might appear that dynamics in the economy are an exception—surely they must contain verbs. But expressed in differential equations, they too are just changes in noun quantities.

Verbs in equation-based theory require workarounds.

## Consequences of restriction to nouns

Let me be clear at this point. I am not saying we shouldn't use mathematics in economic theory. Providing it starts with realistic assumptions, algebraic mathematics allows economists to be precise about the logic of international trade, finance theory, antitrust policy, and central bank policy, and this gives modern theory its power. But it also places certain limitations on what theory can express.

Because algebraic mathematics allows only quantifiable nouns and disbars verbs, it acts as a sieve. What it can't express it can't contain, so processes and actions fall through the sieve and are unexpressed.[6] This noun-restriction causes distortions to the story economics tells. Let me list some of these.

1. **Anything to do with formation or process has to be left out.** If we reduce our discourse to nouns—to objects—it is almost impossible to talk about the actions that take place in a formation process. To give just one example, how economic development happens is not well understood in standard theory. In a noun-world, development can only be the quantitative change in something, and so we build dynamic general-equilibrium models of development: rises and falls of levels of goods produced and their prices. But real development is far more complicated. It consists of people—farmers, traders, small trades people, local functionaries, government officials, financial people—setting up trading arrangements, planting new crops, lending, taking out small loans, opening schools, starting small firms, trying new techniques, failing, buying a new truck here and setting up a delivery route there; as well as institutions forming, morphing, and evolving. It proceeds from millions of actions—heterogenous actions—happening and reinforcing each other. If loans are made, enterprises get set up; if enterprises get set up, people earn money; if money is available for spending, further enterprises get set up. Development is process, a complex set of interacting processes. When I studied the subject in graduate school, "development" was defined via a simple noun criterion: if a country's GDP in growing more than 1% it is developing. This is like reducing the biological development of a new fetus to a criterion of whether its volume is increasing by 1% in a given time. This would yield no science of embryology, and it would give a false illusion that because we can measure something we call "development" we understand the process.

2. **Actions become hidden within unspecified linkages.** What takes the place of actions in economic theory are linkages between economic objects: A decrease in unemployment increases inflation. Investment produces growth. A fall in interest rates induces investment. These statements may be valid, but the story behind them—the detailed actions by which they happen—are not spelled out. The linkages are handled by functions. One key one is the production function: it links inputs to output in manufacturing or services. A steel factory uses iron ore, coke, heavy machinery, and human labor to produce rolled steel; but the activities by which this happens are considered "technology," a kind of "that-which-lies-between" inputs and outputs, not specified and not particularly understood but subsumed mysteriously into a production function.

When actual processes are left out as in this case, they become unseen. And therefore they are not talked about. In 1957 Robert Solow showed that 80% of economic growth—growth of the GDP object—could not be explained by increases in the capital and labor objects. He postulated that the missing factor was "technology," which turned out to be correct. But in the subsequent decades, where technology comes from, how it forms, how it works its way into the economy, were not well explained. These are all processes and couldn't be handled by equations. Take invention itself, the source that feeds this well of economic growth. It is a process or series of processes: of identifying a problem, constructing solutions to it by combining other already existing solutions or developing new ones, testing and experimenting with these, and solving the subproblems these bring up (Arthur, 2007). Notice the verbs. Noun restriction forces economics to remain ignorant of these fundamental actions by which 80% of the economy grows and changes.

---

[6] Some parts of economic theory do allow actions: game theory allows moves and actions. And Austrian economics is based on economic processes. But I am talking about equation-based theory here, which is built from and reverts always to quantifiable objects—nouns and only nouns.

3. **Nouns become idealized conceptual objects.** If we build economic theory on nouns we want them to be general so they can be used across the field. So they become conceptual objects, idealized generalities, not things that are real and specific and tangible. Thus "the Firm" is an ideal object, which brings an artificial homogeneity that doesn't belong, and an applicability that is shaky. What exactly is a firm in this day of platform services and cross-national conglomerates? And so the exactness of nouns becomes an illusion. In much the same way, industries are treated as a conceptual ideal, and to a large extent alike. This uniformity may seem to be a small price for logical precision, but in reality the lack of specificity is a source of imprecision. For instance it makes structural change hard to talk about for theory. Structural change happens when new industries or new technologies change the character of the economy or its parts, as happened when the agrarian economy gave way to the manufacturing one. Theory can't pick up a change in character if its objects of interest don't change in character.

The result is a world of idealized, frozen concepts, one that in another context neuro-philosopher Iain McGilchrist (2009) describes as disembodied, abstracted, isolated from context, and in which "whatever depends on the implicit or can't be brought into focus and fixed, ceases to exist."

4. **Equations bias economics toward equilibrium thinking**. The reason here is simple. Noun-based economics links nouns to other nouns via systems of relation and balance—equations. It is easier to analyze these if they hold still, so to speak, much as it is easier to study the workings of a butterfly if we nail it to a board. And so we purchase understanding by assuming stasis—equilibrium. But the results are at best mixed; all too often the system hangs lifeless, unchanging in time.

5. **Novel creations can't easily emerge.** Once you have an equation system with its x's, y's and z's, it is hard for the system to endogenously generate new variables v's and w's. So novel products, novel strategies, novel ways to use the system, if they are not already declared as variables ready to exist, can't easily emerge. The universe of discussion becomes closed.

6. **Equations bias economics toward rationality.** The reason for this is indirect. In real life, many economic problems—I would say the majority—are not well defined. The players don't quite know what situation they are dealing with or who their competitors will be or what strategies will be on hand. They are subject to fundamental uncertainty, so they can't well-define the "problem" they are facing. It would appear then theoretically they can't move. Yet in the actual economy players deal with unspecified situations all the time, and always via some process. They "make sense" of the situation they are in, and this requires them to explore, try out different strategies, figure what works and what doesn't. But these are processes and can't be handled by nouns or a static system, so equation-based economics is forced to reject this type of situation and assume the problems it considers are well-defined. Agents then choose the best solution available to them; and to make this cognitively plausible for a group of agents, we assume they must do this knowing other agents are like them and will do the same as they do. And so analysis assumes identical actors facing well-defined problems and resolving them identically with unique optimal solutions. It assumes in other words "rational" actors.

§

Many of the above biases have been pointed to before, although from different viewpoints (Simpson, 2013; Cassidy, 2009). What I'm pointing out here is that they derive roundly from algebraic mathematics' restriction to nouns. The most important bias is the choice of what entities to look at in the first place: noun ones, rather than action ones.

The results of this noun-restriction are not uniform across economics. There are two large questions in economic theory: questions of allocation—how quantities of goods and levels of prices are determined within and across markets or trading regions; and questions of formation—how an economy emerges in the first place, how innovation works, how economic development takes place, how structural change happens (Arthur, 2015b). The first by definition deals with quantities and therefore can be expressed in algebraic form. And so equation-based theory has been useful in dealing with questions of allocation (although many heterogeneities, details, and processes may be glossed over in the mathematics). Formation on the other hand by definition has to do with actions, and these cannot easily be reduced to equations. Thus equation-based economic theory has little to say about questions of formation and it has to leave that out. Anything in the economy that deals with adjustment—whether it is adaptation, innovation, structural change, or history itself—falls through the algebraic mathematics sieve. It falls into the domain of literary or case-based description and remains there, partly unspoken.[7]

## II

## Including verbs in economic theory

We have seen that algebraic mathematics may articulate economic theory, but at the cost of severe limitations, so we could stop at this point. But this begs the question: Can we construct economic theory that includes verbs (processes) as well as nouns?

To begin with, let me note that a body of theory with verbs already exists in economics—Austrian economics. It is built very much on process: on individual consumers and producers choosing desired ends, figuring plans to achieve these, changing these on a trial-and-error basis, and discovering new opportunities for profit (Boettke and Coyne, 2015). Because of its process base it doesn't lend itself to algebraic mathematics and is expressed mostly in narrative prose. But this protects it from many of the distortions I have just outlined. It can allow nonequilibrium, non-rationality, and actions, and this gives it a richness we don't find in the standard theory. I believe the Austrian approach deserves a more central place in economic theory.

But I would still ask if we can find a formal—logic-based—approach that includes processes.

One clue is that there exists a formal logic within general mathematics—Boolean logic—that does allow for events. Boolean logic is expressed as statements about the logical implication of truths: If $A$ and $B$ are true or if $K$ is true, then $G$ is true. It is a short step from this to an applied version that says: If process (instructions for) $A$ and process $B$ have been completed, or if process $K$ has been executed, then execute process $G$. Note the if-then conditions directing actions. This is beginning to sound distinctly computational—or algorithmic—so let us hold this in mind.

Another clue is that actions do occur in a core part of economic theory, market-design theory, although they can't be expressed in equation form. Consider the 1962 Gale and Shapley college admissions problem. Students privately rank colleges, and colleges rank students that apply. Students first apply to their preferred set of colleges, and colleges choose their quota of most preferred students, put these on a waiting list, and reject the rest. Rejected students now apply to other colleges, and if preferred by these may replace other students on their wait lists, causing new rejects. Eventually the process terminates when every student is either waited-listed or rejected by all colleges, and Gale and Shapley show it results in a stable matching where no two students would want to swap colleges that would want to swap them. Notice the problem is squarely an economic one, it is an allocation problem, but the solution is a process—an algorithm. The reasoning is also a type of mathematics, logically if not algebraically based. As Gale and Shapley put it, "The argument is carried out not in mathematical symbols but in ordinary English. …Yet any mathematician will immediately recognize the argument as mathematical." The same holds for other market-design systems. Alvin Roth's (2004) bidding system for kidneys and other auction systems are also step-by-step processes. These are algorithmic, yet are part of core economic theory and "mathematical."

---

[7] Sometimes formation, say of a galaxy in physics, can be expressed in equation form, particularly when it deals with multiple elements whose position or momentum or other attributes can be described as quantifiable nouns.

*Algorithms as Carriers of Process*

Both these clues hint strongly that algorithms, or computation if you prefer, might be a way to include processes. Indeed, computational studies have been present in economics for quite some time (see Roth, 2001; Mirowski, 2002; Tesfatsion, 2006) and are steadily growing in number.

---

**Box 1**: Can algorithms constitute "theory"?

To answer this let me borrow some insights from a recent branch of mathematics called algorithmic information theory (AIT), developed by Gregory Chaitin (1990, 2006, 2012) and others. The key insight of AIT is that we can think of an "algorithm" as a set of core specifications or instructions for a subsequent unfolding—a subsequent "computing out"—of these instructions. If we have an algorithm that says write two 0s, then two 1s, then repeat, it will compute out to 0011 0011 0011 0011 indefinitely. The algorithm is a *compression* of the information of the sequence of 0011 0011 0011 …. And the computed-out sequence is a *decompression* of the algorithm's concise instructions. In this way a score in music is an algorithm, a compression of information, and the performed symphony its decompression or unfolding of the score's information.

This surprisingly is where algorithms link with standard mathematics. Differential equations perform much the same role as algorithms in standard mathematics; they are sets of compressed instructions for a subsequent unfolding or integrating out. (Similarly, we can think of static equations as core specifications and their "unfolding" as the implications they carry.) This view is useful because it shows us that algorithms are not so different from equations. Both systems are compressed information that can be unfolded to arrive at their logical consequences. Both can express situations in science, both can be probed—queried, unfolded—for their implications, and both are modes of expression: one system allows only quantifiable nouns; the other allows both nouns and verbs.

Both systems—algorithms as well as equations—can also constitute theory. Chaitin points out that theories are condensed descriptions of how some system works. Newton's equations are condensed descriptions of the motions of planets: we can decompress the information contained in them to arrive at elliptical planetary orbits and as such his system constitutes a theory of planetary orbital information. The same can be said for algorithms. Our algorithm that produces 0011 0011 indefinitely is a compression of the information of such a system, so just as with Newton we can say it constitutes a theory of the system. Algorithms used to describe systems—to condense the information in a complicated situation—are theories of such systems. If economic models based on algorithmic specifications are condensed descriptions of how systems work, we can see them as theory proper, just as we see equation-based descriptions as theory proper.

---

The word "computation" has strong associations with computers, which are not always necessary in what follows, so I will avoid that term and talk instead of "algorithms." This label isn't perfect either (it has popular associations with AI or with social media marketing), but algorithms are established entities in mathematics, are based firmly on Boolean logic, and allow moves or actions. So let us look more deeply at how algorithmic expression could be used to construct an economics that allows processes.

Algorithms, of course, are sets of instructions or operations that are executed by some system. They are by no means new—in the form of calculation methods they go back to Babylonian times. They exist as entities in themselves, and they do not necessarily require a physical computer. We can think of the replication steps of a DNA strand as an "algorithm," and the replication process as the outcome of the steps' instructions. Algorithms can allow for randomness, can operate in continuous time, and can have several threads operating in parallel. They can also constitute "theory" (Box 1).

Algorithms of course can allow noun-objects, but these don't need to be quantifiable: they can deal with customer names as readily as dollar numbers. And they also allow verbs or processes actions.

The reason algorithms handle processes well is because each individual instruction, each step, can signal an action or process. Also, and here is where process enters par excellence, they allow *if-then* conditions. *If* process R has been executed 100 times, *then* execute process L; if not, *then* execute process H. Algorithms can contain processes that call or trigger other processes, inhibit other processes, are nested within processes, indeed create other processes. And so they provide a natural language for processes, much as algebra provides a natural language for noun-quantities. Frequently algorithms include equations, and so sometimes we can think of algorithmic systems as equation-based ones with if-then conditions. As such, algorithmic systems generalize equation-based ones, and they give us a new mode, a new language of expression in economics, although one that may look different from what we're used to.

What about the assertion that Gale and Shapley made that algorithms constitute a form of mathematics? Mathematician Gregory Chaitin (2012) argues that just as we regard a set of differential equations along with their initial values as a mathematical object, and integrating these out—unfolding their consequence—as a mathematical operation, so we can regard the core instructions plus initial data of an algorithm as a mathematical object, and computing these out—unfolding their consequence—as a mathematical operation. Both are mathematics, albeit expressed in different symbolic languages. "The computer," says Chaitin (2012), "is a revolutionary new kind of mathematics with profound philosophical consequences. It reveals a new world." Indeed, in 1936 Alan Turing had viewed "methods"—operations performed on data and producing calculated outcomes, which we'd now call algorithms—as valid objects within mathematics to be manipulated logically and used in proofs. To him they revealed a new world. Edsgar Dijkstra expressed much the same thought when he observed that "Computer science is no more about computers than astronomy is about telescopes."

*Reservations, and Some Comments*

The reader may feel at this stage some misgivings about admitting algorithms—or computation—to the hallowed halls of economic theory, so let me address these for a moment.

One misgiving is that equation systems are exact, but algorithms are not. This sounds plausible, but actually it isn't. The exactness of equations is an illusion. Equations as I said use noun-objects that might have been put together from some idealized form, and linked to other noun-objects by coefficients or rate parameters. "As soon as you write an equation it is wrong," said physicist Albert Tarantola, "because reducing a complex reality to an equation is just too simplistic a view of things" (quoted in Bailey 1996, p.29). In fact, much of economic theory consists of pointing out how averaged things (unemployment, say) affect averaged things (inflation, say), which yields at best coarse-grained theory. Algorithms can avoid such mean-field approximations. They can include proper detail and proper heterogeneity, and where these matter algorithms are more accurate—and more rigorous.

A second misgiving is that algorithms are "simulations" and as such are often built on *ad-hoc* assumptions. Certainly in the wrong hands algorithms can be subject to chicanery, but so too can equation systems as Paul Romer (2015) has pointed out. What counts with either mode of expression is rigor in the form of honesty and exactness in modeling.

Using algebraic mathematics or algorithmic mathematics in economics has different advantages and drawbacks. Both systems trace a pathway from agent behavior to its implied outcome. With algebraic mathematics we can prove things exactly and reliably; we can follow the logic step by step and understand what it reveals. Algorithms lack this advantage, but they compensate in other directions. We can often regard algorithmic computations as laboratory experiments, running them to see what they reveal and exploring what phenomena appear and the conditions under which they appear. We can free ourselves from the noun-restrictions I listed earlier and allow proper process. We can expand algorithms to encompass an arbitrary amount of realistic detail. And we can see their *if–then* conditions as allowing the changing context of the situation—the 'if' clause can identify where the computation currently is, the external or internal circumstances that matter—to direct agents' behavior in any way appropriately called for (Arthur, 2020). Agents can react even in ill-defined situations with some modicum of intelligence.

Even with these advantages, I do not believe algorithmic expression is a panacea in economics. It can include heterogenous agents with "non-rational" behavior that is context dependent, detailed and therefore more realistic. But it does not easily capture the "humanness" of economic life, its emotionality, its intuitive nature, its personages, its very style. For this we would need other means.

**What sort of world gets revealed?**

I said earlier that methods in science can be instruments of seeing or of understanding, and act to reveal different things. What sort of things—what sort of world—would economic reasoning expressed in algorithmic terms reveal?

Certainly, by allowing processes to be included, we would get a much more direct understanding of economic development, of how institutions form, of how technological innovation works and transforms the economy, and we could thus bring these into formal theory. And because processes are always in dynamic flux (Dupré, 2017), we would be aware that the entities they are operating on are open and changing, and certainly not in equilibrium. We might see also that simple processes can lead to complex and unpredictable outcomes (Wolfram, 2002).

What would economics look like overall if it were a more procedural—more verb-based—science? Biology gives us an idea. It is a procedural discipline, not based on quantities growing and changing, but rather on processes: processes that determine the step-by-step formation of structures (embryology); processes that respond to their internal and external environment (immune response, gene expression[8], neural responses); processes that create novelty (speciation and adaptation). Nested within these processes are further processes, and within these still further processes, all in a multilayer hierarchy with processes triggering each other or inhibiting each other possibly randomly in complicated networks of interaction. "In the living world," says Dupré (2017), "a metaphysics of 'things' is hard to sustain … all we are left with are highly dynamic processes."[9]

The natural language of life is algorithmic.

By this token, admitting process to economics would reveal a world where structures large and small continually form, where agents and organizations continually respond to their internal and external environment and change from within as they do, where fresh undertakings continually create novelty. The economy would become a living thing.

Such a view chimes with the new approach of complexity economics (see Arthur, 1999, 2015a, 2021).[10] This sees agent behavior not as *consistent with* (in equilibrium with) the outcome it brings about, but as *reacting to* the outcome it causes. In this economics, agents in the economy—producers, investors, banks—differ in ways unknown, and thereby inhabit a world where fundamental uncertainty is the norm. They act and react to the changing circumstances they mutually create: they imagine, attempt to "make sense," test out ideas, and occasionally come up with novel responses. These reactions in turn change the situation or outcome, so that agents may have to react or adjust again. No equilibrium need necessarily emerge. We are in a world of constant change and response, one not of stasis but of process. All economies are systems of processes permanently in formation.

---

[8] See Isabelle Peter and Eric Davidson (2015).

[9] For a process-based philosophy of biology see Nicholson and Dupré (2018). Rigorous philosophy based on process goes back to A. N. Whitehead (1927). For modern versions see Rescher (2000), Kaaronen (2018) and Hertz, Mancilla-Garcia, and Schlüter (2020).

[10] A closely related approach based is agent-based computational economics (see Tesfatsion and Judd, 2006; Axtell and Farmer, 2021). The two approaches exist on a continuum; both allow algorithmic modeling. Complexity economics emphasizes economic theory; agent-based computational economics as its name suggests emphasizes computation.

The world revealed here is not one of rational perfection, nor is it mechanistic. If anything it looks distinctly biological. Its agents are constantly acting and reacting within a situation—an "ecology" if you like—brought about by other agents acting and reacting. Algorithmic expression allows novel, unthought of behaviors, novel formations, structural change from within—it allows creation. It gives us a world alive, constantly creating and re-creating itself.

## Conclusion

All theories in economics are constructions of thought about human behavior using some instrument or mode of understanding, and I don't believe there is a final, ideal mode that can reveal all there is to know about the economy. Whether the mode of understanding is close observation expressed discursively, historical analysis, geometrical reasoning, algebraic mathematics, or algorithmic expression, all are valid modes even if they apply best to different contexts, and all reveal different things.

Algebraic mathematics is a particularly powerful instrument for economic theory and it will and should continue to have a valid place in economics. It reveals the implications of the large equilibrating forces that hold the economy together, and the logic that governs its smaller structures. But it comes with the requirement that its explanations—the "stories" it tells—must be put together only with nouns, quantitative nouns. This requisite is so familiar we do not even notice it; in fact if we are adept in mathematics we feel comfortable with it. But that doesn't diminish its consummate awkwardness—think of trying to write a novel using only nouns.

Algorithmic mathematics is different. It allows verbs and thereby unleashes actions—processes and processes within processes. It allows contingency, the idea that circumstances—perhaps randomly arrived at—determine what happens next. It allows context: agents respond to detailed aspects of the overall outcome they together create. It provides a way of exploring the formation of structures, the ongoing process of nonequilibrium, and indeed how history creates the economy and the economy creates history.

As a means of understanding, algorithmic expression need not replace equation-based expression in economics, but can take its place alongside it as a parallel language, as equations once did for geometry.

## Acknowledgments

I am grateful to Ronan Arthur, James Bailey, Richard Bronk, David Colander, Doug Erwin, Kristian Lindgren, Deirdre McCloskey, Iain McGilchrist, Rika Preiser and David Simpson for comments. All views expressed above are strictly my own.